\documentclass[journal=jacsat,manuscript=article]{achemso}

\usepackage{chemformula} 
\usepackage[T1]{fontenc} 
\usepackage{graphicx}
\usepackage{soul} 
\usepackage{bm}
\usepackage{amsfonts}
\graphicspath{
{./figure/}
{./}
}

\author{Kairi Masuda}
\email{kairi.masuda.c5@tohoku.ac.jp}
\author{Yu Kumagai}

\affiliation[Tohoku University]
{Institute for Materials Research, Tohoku University, 2-1-1 Katahira, Aoba-ku, Sendai, 980-8577, Japan}

\title[An \textsf{achemso} demo]
{Atomic-scale phase-field modeling with dopants:\\ Stochastic self-consistent harmonic approximation with fractional site occupancy}

\abbreviations{IR,NMR,UV}
\keywords{American Chemical Society, \LaTeX}

\begin{document}







\begin{abstract}
Phase-field modeling has achieved great success in predicting pattern formation in materials, such as the formation of ferroelectric domains.
However, because it is typically based on continuum mechanics, conventional phase-field modeling cannot be straightforwardly applied to atomic-scale pattern formation, such as dopant segregation and vacancy ordering, which are driven by chemical potentials.
Here, we extend the phase-field concept to the atomic scale by formulating the free energy of atomic systems within stochastic self-consistent harmonic approximation (SSCHA) theory to allow fractional site occupations.
Our methodology enables us to directly calculate the derivative of the free energy with respect to site occupation and thereby obtain the chemical potential, successfully reproducing Ag distributions in bulk Cu as well as the resulting lattice expansion.
Furthermore, we applied our methodology to investigate dopant segregation around a $\Sigma5(310)[001]$ Cu grain boundary doped with Ag atoms.
We found that Ag atoms preferentially segregate at the vertices of the triangular motif of a grain boundary.
As the number of dopants increases, excess Ag atoms segregate near the vertices and then at the bottom sites of the triangular motif.
This study extends the phase-field concept to discrete atomic systems, enabling the identification of preferential dopant-segregation sites and thereby visualizing atomic-scale pattern formation.
\end{abstract}

\clearpage

\section{Introduction}
In phase-field modeling\mbox{\cite{LQchenReview,SolidificationReview,Ambati2015,doi:10.1098/rsta.2015.0166}}, phenomena of interest are described in terms of one or more field variables. The associated free energy is expressed as a spatially dependent functional of these fields\mbox{\cite{doi:10.1080/14786444908561372,doi:10.1080/14786445108561354}}. This local thermodynamic description has achieved great success in reproducing a wide variety of pattern-formation phenomena in materials science, such as polar vortices and skyrmions in nanoscale ferroelectric systems\mbox{\cite{LQchen_review_thin_film,doi:10.1038/nature16463,doi:10.1038/s41586-019-1092-8,doi:10.1038/s41563-020-0694-8,doi:10.1038/s41467-023-36950-x,doi:10.1126/science.1259869}}. In addition to describing such spatially nonuniform patterns, phase-field simulations can also provide position-resolved thermodynamic quantities, including local free-energy and entropy distributions\mbox{\cite{CHOUDHURY20055313,Gao2013,Park2018,WANG2021117383,NatureYang2025,doi:10.1126/science.abb3209}}.
In this sense, phase-field modeling functions as a computational microscope that visualizes pattern formation and the underlying local thermodynamic states. Meanwhile, recent advances in materials science have increasingly highlighted phenomena at much smaller length scales, where structural and thermodynamic heterogeneity can extend down to the scale of individual atoms\mbox{\cite{doi:10.1126/science.adq4147,doi:10.1038/s41586-020-2082-6,PhysRevLett.120.267601,doi:10.1126/science.adh7670,Horiuchi2008,elastin,doi:0.1038/s41467-021-21019-4,doi:10.1038/36069}}. However, because conventional phase-field and phase-field-crystal models are generally formulated at continuum or crystalline length scales\mbox{\cite{doi:10.1103/PhysRevE.70.051605,doi:10.1103/PhysRevE.79.035701,doi:10.1103/PhysRevE.73.031609,doi:10.1007/s11837-007-0095-3,doi:10.1103/PhysRevMaterials.4.013802}}, they cannot be straightforwardly applied to systems in which the relevant heterogeneity is intrinsically atomistic. These developments therefore point to the need for a local thermodynamic framework capable of resolving thermodynamic states at the single-atom level.

A further enhancement of thermodynamic resolution can be achieved by regarding the probability distribution of atomic vibrations itself as a field variable. In this formulation, a variational free-energy functional is constructed from this atomic field and an interatomic potential\mbox{\cite{PhysRevLett.63.624,doi:10.1063/1.460547,PhysRevB.84.054103,PhysRevB.110.104107,doi:10.1103/ngky-crt4}}, and the distribution widths or underlying force constants are optimized to minimize the free energy. This self-consistent harmonic approximation (SCHA)-type theory has been further extended to the stochastic self-consistent harmonic approximation (SSCHA)\mbox{\cite{PhysRevB.89.064302,PhysRevB.98.024106,doi:10.1038/s41586-020-1955-z,doi:10.1088/1361-648X/ac066b,PhysRevB.110.144101}}, in which variational free energies are stochastically evaluated using forces and energies obtained from first-principles calculations or many-body machine-learning interatomic potentials for complex materials. Recently, (S)SCHA-type theories have been applied not only to calculate free energies as global quantities but also to evaluate local free-energy distributions and related quantities, such as stress and entropy. Thus, (S)SCHA theory has the potential to be used for studying pattern formation and visualizing local thermodynamic states with atomic resolution, namely, atomic-scale phase-field modeling.

One of the next steps in developing atomistic phase-field modeling is to calculate local chemical potentials and use them to describe related atomic-scale pattern-formation phenomena, such as dopant segregation, vacancy ordering, and inhomogeneous elemental distributions in high-entropy materials, all of which can drastically alter material properties\mbox{\cite{doi:10.1038/ncomms11079,doi:10.1038/s41467-024-49437-0,doi:10.1038/s41467-022-30018-y}}. In conventional molecular simulations, however, calculating chemical potentials from the free energy is difficult because site occupancies are discrete variables and are therefore not differentiable. Li et al. proposed incorporating fractional site occupancy into an SCHA-type free energy\mbox{\cite{PhysRevB.84.054103}}, which makes the free energy differentiable with respect to site occupancy and thus enables the evaluation of chemical potentials. They showed that this approach can be used to calculate vacancy diffusion, thereby demonstrating its applicability to diffusive molecular dynamics simulations. However, the formalism of Li et al. is limited to pair potentials and isotropic Gaussians, which restricts its applicability and accuracy. Therefore, a free-energy functional that incorporates both fractional site occupancy and many-body interactions is needed to further explore pattern formation at the atomic scale.

In this work, we develop an atomic-scale phase-field modeling framework for dopant-containing systems by combining SSCHA theory with fractional site occupancy. This framework enables the calculation of site-resolved dopant-segregation patterns at the atomic scale. The remainder of this paper is organized as follows. First, we describe the theoretical formulation for incorporating fractional site occupancy into the SSCHA free-energy framework. Second, we apply the developed methodology to Ag-doped bulk Cu and calculate the Ag distribution and the associated lattice expansion to validate the proposed framework. We then apply the methodology to Cu grain boundaries and visualize dopant-segregation patterns. Finally, we summarize our findings and discuss future directions for this modeling approach.

\clearpage

\clearpage
\section{Methods}

\subsection{Theory}
\subsubsection{Free energy of an Einstein solid}
An Einstein solid is described by the following harmonic potential:
\begin{eqnarray}
U(\bm{x})=\frac{1}{2}k_{ij}x_{i}x_{j},
\end{eqnarray}
where $k_{ij}$ is the force-constant matrix and $\bm{x}$ is the displacement from the mean position. 
The thermal vibration of an atom around its mean position appears as a probability-density cloud over a longer time scale; that is, $\rho \propto \exp(-U/k_{B}T)$, where $k_{B}$ is the Boltzmann constant and $T$ is the temperature. These probability-density clouds for Einstein solids follow Gaussian distributions\mbox{\cite{PhysRevLett.63.624,doi:10.1063/1.460547,PhysRevB.84.054103,PhysRevB.110.104107}}, whose general form can be described as follows:
\begin{eqnarray}
\rho(\boldsymbol{r}|\boldsymbol{X}, \boldsymbol{\Sigma})&=&
\frac{1}{(2\pi)^{3/2}\sqrt{\mathrm{det}\boldsymbol{\Sigma}}}\\ \nonumber
&\times&\exp  \Big\lbrace -\frac{1}{2}(\boldsymbol{r}-\boldsymbol{X})^T 
\boldsymbol{\Sigma}^{-1}(\boldsymbol{r}-\boldsymbol{X}) \Big\rbrace.
\end{eqnarray}
Here, $\bm{x} = \boldsymbol{r} -\boldsymbol{X}$, where $\boldsymbol{r}$ is the position of the oscillator and $\boldsymbol{X}$ is its mean position.
$\mathrm{det}\boldsymbol{\Sigma}$ is the determinant of the covariance matrix, and $\boldsymbol{\Sigma}^{-1}$ is its inverse. The covariance matrix is represented as follows:
\begin{eqnarray}
\boldsymbol{\Sigma}=\bm{R}\begin{pmatrix}
\sigma_{x'}^{2} & 0  & 0 \\
0 & \sigma_{y'}^{2}  &  0  \\
0 & 0 & \sigma_{z'}^{2}
\end{pmatrix}\bm{R}^{T},
\end{eqnarray}
where $\bm{R}$ is a three-dimensional rotation matrix, which can be decomposed into rotation matrices about each axis as follows:
\begin{eqnarray}
\bm{R}= \bm{R}_{z}(\theta_{z})\bm{R}_{y}(\theta_{y})\bm{R}_{x}(\theta_{x}).
\end{eqnarray}
Therefore, the Gaussian distribution is specified by six variables, namely, $\sigma_{x'}$, $\sigma_{y'}$, $\sigma_{z'}$, $\theta_{x}$, $\theta_{y}$, and $\theta_{z}$, where $\sigma_{x'}$, $\sigma_{y'}$, and $\sigma_{z'}$ are the standard deviations along the principal axes, and $\theta_{x}$, $\theta_{y}$, and $\theta_{z}$ are the rotation angles around each axis. When $\boldsymbol{X}$ and $\boldsymbol{\Sigma}$ are given, the probability density at position $\boldsymbol{r}$ is determined. Using these Gaussian shape parameters, the free energy of the Einstein oscillator is represented as follows:
\begin{eqnarray}
F_{\rm{Einstein}}&=&\frac{1}{2}k_{B}T 
 \ln \left(\frac{\Lambda^{6}}{8\pi^{3} 
\sigma_{x'}^2 \sigma_{y'}^2 \sigma_{z'}^2} \right),
\end{eqnarray}
where $\Lambda=\hbar\sqrt{2\pi/(m k_{B}T)}$ is the thermal de Broglie wavelength, $\hbar$ is the reduced Planck constant, and $m$ is the mass of the atom.

\subsubsection{Variational free energy with site-occupancy probabilities}
The Gibbs--Bogoliubov inequality provides an upper bound on the true free energy, $F_{\rm{true}}$, as follows:
\begin{eqnarray}
F_{\rm{true}}\leq F_{0}+\langle \phi-V\rangle_{0} \equiv F,
\label{eq:GibbsBogo}
\end{eqnarray}
where $F_{\rm{true}}$ is the true free energy associated with the interatomic potential $\phi$, while $F_{0}$ is the free energy of the reference system associated with the reference potential energy $V$. The notation $\langle \cdot \rangle_{0}$ denotes the expectation value with respect to the reference probability distribution governed by $V$. 
LeSar \textit{et al.} employed an Einstein solid as the reference state, i.e., $F_{0}=F_{\rm{Einstein}}$ and $V=\sum U$, and derived an analytical form of $F$ using mutually independent isotropic Gaussians\mbox{\cite{PhysRevLett.63.624,doi:10.1063/1.460547,PhysRevB.84.054103}} for Cu described by a Morse potential. Here, we extend their theory to general Gaussian distributions\mbox{\cite{PhysRevB.110.104107}} and further generalize it to many-body interactions as follows:
\begin{eqnarray}
\label{eq:freeene}
F&=&\frac{1}{2}k_{B}T\sum_{i=1}^{N} 
\left \{ \ln \left(\frac{\Lambda_{i}^{6}}{8\pi^{3} 
\sigma_{i,x'}^2 \sigma_{i,y'}^2 \sigma_{i,z'}^2} \right)-3 \right\} \\
&+&\int\int\cdots\int
{\rho_{1}(\boldsymbol{r}_{1})}{\rho_{2}(\boldsymbol{r}_{2})}\cdots
{\rho_{N}(\boldsymbol{r}_{N})}
\phi(\boldsymbol{r}_{1},\boldsymbol{r}_{2},\ldots,\boldsymbol{r}_{N})
d\boldsymbol{r}_{1}d\boldsymbol{r}_{2}\cdots d\boldsymbol{r}_{N}, \nonumber
\end{eqnarray}
where $N$ is the number of atoms. The first term is the free-energy contribution from atomic vibrations, which corresponds to $F_{0}-\langle V \rangle_{0}$ in Eq.~(\ref{eq:GibbsBogo}). Note that $\langle \sum U \rangle_{0} = \frac{3}{2}Nk_{B}T$ according to the equipartition theorem. The second term is the expectation value of the many-body potential energy $\phi$, i.e., an effective potential among the Gaussian clouds, and corresponds to $\langle \phi \rangle_{0}$. For given mean positions $\boldsymbol{X}_{i}$, the estimation of $F_{\rm{true}}$ is obtained by optimizing the Gaussian shapes $\boldsymbol{\Sigma}_{i}$ so as to minimize the upper-bound free energy $F$.

To extend the above many-body free energy to a semi-grand-canonical formulation, we introduce a continuous occupation probability $c_{i}$, where $c_{i} = 0$ indicates that site $i$ is occupied by a host atom, whereas $c_{i} = 1$ indicates that the site is occupied by a dopant atom. The free energy is then written as follows (the derivation is given in Supplemental Information S1):
\begin{eqnarray}
F&=&\frac{1}{2}k_{B}T\sum_{i=1}^{N}\Bigg[ (1-c_{i}) \left \{ \ln \left(\frac{\Lambda_{i,\rm{host}}^{6}}{8\pi^{3} \sigma_{i,x'}^2 \sigma_{i,y'}^2 \sigma_{i,z'}^2} \right)-3 \right\} \\ \nonumber
&+&c_{i} \left \{ \ln \left(\frac{\Lambda_{i,\rm{dopant}}^{6}}{8\pi^{3} \sigma_{i,x'}^2 \sigma_{i,y'}^2 \sigma_{i,z'}^2} \right)-3 \right\}\Bigg]  \\ \nonumber
\\[0.1 ex] \nonumber
&+& \sum_{\bm{n}\in\{0,1\}^{N}} c_{1}^{n_{1}}(1-c_{1})^{1-n_{1}}
c_{2}^{n_{2}}(1-c_{2})^{1-n_{2}}\cdots
c_{N}^{n_{N}}(1-c_{N})^{1-n_{N}} \\ \nonumber
&\times&\int\int\cdots\int
{\rho_{1}(\boldsymbol{r}_{1})}{\rho_{2}(\boldsymbol{r}_{2})}\cdots
{\rho_{N}(\boldsymbol{r}_{N})}
\phi(n_{1},n_{2},\ldots,n_{N},\boldsymbol{r}_{1},\boldsymbol{r}_{2},\ldots,\boldsymbol{r}_{N})
d\boldsymbol{r}_{1}d\boldsymbol{r}_{2}\cdots d\boldsymbol{r}_{N}\\ \nonumber
\\[0.1 ex] \nonumber
&+&k_{B}T \sum_{i=1}^{N}[(1-c_{i})\ln (1-c_{i})+ c_{i} \ln c_{i}],
\end{eqnarray}
where $\Lambda_{i,\rm{host}}$ and $\Lambda_{i,\rm{dopant}}$ are the thermal de Broglie wavelengths of the host and dopant atoms at site $i$, respectively, and $n_{i}$ is a binary indicator variable defined as follows:
\begin{equation}
  n_{i}=
  \begin{cases}
    1 & \text{if site $i$ is occupied by a dopant atom,} \\
    0 & \text{if site $i$ is occupied by a host atom.}
  \end{cases}
\end{equation}

\subsubsection{Governing equations}
For the given positions $\boldsymbol{X}_{i}$ and the optimal parameters $\boldsymbol{\Sigma}_{i}$ calculated above, the atomic probability densities should evolve in time so as to reduce the free energy while conserving the normalization condition, i.e., $\int \rho_{i} \, d\boldsymbol{r} = 1$. This implies that the probability densities should evolve according to the following conservative phase-field equation:
\begin{eqnarray}
\label{eq:govern}
  \frac{\partial \rho_{i}(\boldsymbol{r},t)}{\partial t}
  &=& \nabla \cdot \left(D_{i}(\boldsymbol{r})
  \nabla \frac{\delta F}{\delta \rho_{i}} \right) \\ \nonumber
  &=& \nabla \cdot \left(\kappa_{i} \rho_{i}(\boldsymbol{r})
  \nabla \Phi_{i}(\boldsymbol{r})\right),
\end{eqnarray}
where $t$ denotes time and $D_{i}(\boldsymbol{r})$ is a position-dependent kinetic coefficient. Here, we set $D_{i}(\boldsymbol{r}) = \kappa_{i} \rho_{i}(\boldsymbol{r})$, where $\kappa_{i}$ is a kinetic coefficient, so that the flux is proportional to the probability density and the normalization of $\rho_{i}$ is conserved. The gradient of the functional derivative, $\nabla (\delta F/\delta \rho_{i}) = \nabla \Phi_{i}(\boldsymbol{r})$, acts as the thermodynamic driving force. By solving the above governing equation, the time evolution of the probability densities can be calculated.

Simultaneously, the occupation probabilities evolve according to the following equation:
\begin{eqnarray}
\frac{d c_{i}}{dt}=\sum_{j\in N(i)} M_{ij} \left(\mu_{j}-\mu_{i}\right),
\end{eqnarray}
where $N(i)$ denotes the set of neighboring sites around site $i$, $M_{ij}$ is a kinetic coefficient for occupation exchange between sites $i$ and $j$, and $\mu_i$ is the site-resolved chemical potential.

\subsection{Simulation detail}
\subsubsection{Simulation targets}
To validate the proposed theory, we selected bulk Cu doped with Ag atoms and evaluated its pressure evolution in the canonical (NVT) ensemble. 
A $2 \times 2 \times 2$ supercell was constructed based on the conventional unit cell of Cu, which has a lattice constant of 3.61 \mbox{\AA} and contains four atoms. The resulting supercell has lattice constants of 7.22 \mbox{\AA} $\times$ 7.22 \mbox{\AA} $\times$ 7.22 \mbox{\AA} and contains $N = 32$ atoms.

\subsubsection{Implementation details of the free-energy calculation}
Figure 1 presents a schematic overview of the computational workflow. The second term in the free-energy functional in Eq. (8) is a high-dimensional integral that can be efficiently evaluated by Monte Carlo techniques as follows:
\begin{eqnarray}
&&\sum_{\bm{n}\in\{0,1\}^{N}} 
c_{1}^{n_{1}}(1-c_{1})^{1-n_{1}}\cdots
c_{N}^{n_{N}}(1-c_{N})^{1-n_{N}} \\ \nonumber
&\times&\int\cdots\int
{\rho_{1}(\boldsymbol{r}_{1})}\cdots{\rho_{N}(\boldsymbol{r}_{N})}
\phi(n_{1},\ldots,n_{N},\boldsymbol{r}_{1},\ldots,\boldsymbol{r}_{N})
d\boldsymbol{r}_{1}\cdots d\boldsymbol{r}_{N} \\ \nonumber
\\[0.1 ex] \nonumber
&=&\mathbb{E}_{u_k\sim U(0,1),\,\bm{\eta}_k\sim N(\boldsymbol{0},\mathbf{I})}
\Big[
\phi \big(
\bm{1}_{\{u_{1}\leq c_{1}\}},\ldots,\bm{1}_{\{u_{N}\leq c_{N}\}},
\boldsymbol{X}_{1}+\boldsymbol{L}_{1}\bm{\eta}_{1},\ldots,
\boldsymbol{X}_{N}+\boldsymbol{L}_{N}\bm{\eta}_{N}
\big)
\Big],
\end{eqnarray}
where $U(0,1)$ denotes the uniform distribution on $[0,1]$, $N(\boldsymbol{0},\mathbf{I})$ denotes the standard three-dimensional normal distribution. $\mathbb{E}_{u_k,\boldsymbol{\eta}_k}$ denotes the expectation over $u_k\sim U(0,1)$ and $\boldsymbol{\eta}_k\sim \mathcal{N}(\boldsymbol{0},\mathbf{I})$ for $k=1,\ldots,N$. Note that $\boldsymbol{r}_{i}=\boldsymbol{X}_{i}+\boldsymbol{L}_{i}\boldsymbol{\eta}_{i}$, where $\boldsymbol{L}_{i}=\boldsymbol{R}_{i}\text{diag} (\sigma_{i,x'},\sigma_{i,y'},\sigma_{i,z'})$ and displacements $\bm{x}_{i}=\bm{L}_{i}\bm{\eta}_{i}$.  $\bm{1}_{\{u_{i}\leq c_{i}\}}$ is an indicator function defined as follows:
\begin{equation}
  \bm{1}_{\{u_{i}\leq c_{i}\}}=
  \begin{cases}
    \ 1 \quad (\text{site $i$ is occupied by Ag}) & \text{if $u_{i}\leq c_{i}$,} \\
    \ 0 \quad (\text{site $i$ is occupied by Cu}) & \text{if $u_{i}> c_{i}$.}
  \end{cases}
\end{equation}
To reduce the statistical error of the Monte Carlo estimate, we employ Sobol low-discrepancy sequences\mbox{\cite{SOBOL196786}} for both $u_{k}$ and $\bm{\eta}_{k}$, and map the sequence for $\bm{\eta}_{k}$ to the Gaussian ensemble through the inverse cumulative distribution function, resulting in a total sample size of $N_{\rm{sample}}=1000$. The interatomic potential $\phi$ is provided by the Multi-Atomic Cluster Expansion (MACE) model\mbox{\cite{Batatia2025}}, which encodes each atomic configuration as a graph and is trained on $r^2$SCAN data. Because MACE can evaluate multiple atomic graphs concurrently, the integrand is computed in parallel on a GPU via PyTorch\mbox{\cite{paszke2017automatic}}, allowing the high-dimensional integral to be evaluated with minimal wall-time overhead.

The gradient of the free energy with respect to the Gaussian width $\sigma_{i,j'}$ is evaluated as follows (For simplicity, the $x'$-direction case is shown):
\begin{eqnarray}
\frac{\partial F}{\partial \sigma_{i,x'}}
&=&-\frac{k_{B}T}{\sigma_{i,x'}} \\ \nonumber
&+&
\mathbb{E}_{u_k,\boldsymbol{\eta}_k}
\Big[
\nabla_{\boldsymbol{r}_{i}}\phi
\big(
\bm{1}_{\{u_{1}\leq c_{1}\}},\ldots,\bm{1}_{\{u_{N}\leq c_{N}\}},
\boldsymbol{X}_{1}+\bm{L}_{1}\boldsymbol{\eta}_{1},\ldots,
\boldsymbol{X}_{N}+\bm{L}_{N}\boldsymbol{\eta}_{N}
\big)
\cdot
\mathbf{R}_{i}\boldsymbol{e}_{x'}\eta_{i,x'}
\Big],
\end{eqnarray}
where $\boldsymbol{e}_{x'} = (1, 0, 0)^T$ is the unit vector in the principal-axis basis, and $\nabla_{\boldsymbol{r}_{i}}\phi$ denotes the gradient of the potential energy with respect to the position of atom $i$. In the same manner, the gradients with respect to the rotation angles are expressed as follows:
\begin{eqnarray}
\frac{\partial F}{\partial \theta_{i,x}}
&=&
\mathbb{E}_{u_k,\boldsymbol{\eta}_k}
\Bigg[
\nabla_{\boldsymbol{r}_{i}}\phi
\big(
\bm{1}_{\{u_{1}\leq c_{1}\}},\ldots,\bm{1}_{\{u_{N}\leq c_{N}\}},
\boldsymbol{X}_{1}+\bm{L}_{1}\boldsymbol{\eta}_{1},\ldots,
\boldsymbol{X}_{N}+\bm{L}_{N}\boldsymbol{\eta}_{N}
\big) \\ \nonumber
&\times&
\Bigg(
\frac{\partial \mathbf{R}_{i}}{\partial \theta_{i,x}}
\mathrm{diag}(\sigma_{i,x'},\sigma_{i,y'},\sigma_{i,z'})
\boldsymbol{\eta}_{i}
\Bigg)
\Bigg].
\end{eqnarray}
The subsequent minimization of the free-energy functional is carried out using the Limited-Broyden–Fletcher–Goldfarb–Shanno (L-BFGS) algorithm implemented in SciPy\mbox{\cite{LBFGS,2020SciPy-NMeth}}, with a relative convergence tolerance of $1.0 \times 10^{-6}$.

\subsubsection{Implementation details of the governing equations}
The phase-field equation, Eq. (\mbox{\ref{eq:govern}}), has the same form as a drift-type Fokker--Planck equation, written as follows:
\begin{equation}
  \frac{\partial \rho_{i}(\boldsymbol{r},t)}{\partial t}
  =
  \nabla \cdot \left( \boldsymbol{a}_{i}(\boldsymbol{r}) 
  \rho_{i}(\boldsymbol{r},t) \right),
\end{equation}
where $\boldsymbol{a}_{i}(\boldsymbol{r})$ is the drift velocity. 
In the present case, the drift velocity corresponds to $\boldsymbol{a}_{i}(\boldsymbol{r})=\kappa_{i}\nabla \Phi_{i}(\boldsymbol{r})$. 
To simplify this equation, we assume that each Gaussian distribution moves without deformation during an infinitesimal motion, and the Gaussian shape is updated subsequently. Under this assumption, the equation can be converted into the following deterministic Langevin-type equation\mbox{\cite{PhysRev.36.823,PhysRevB.110.104107}}:
\begin{equation}
\label{eq:langevin}
  \frac{d\boldsymbol{X}_{i}}{dt}
  =
  -\kappa_{i} \nabla_{\boldsymbol{X}_{i}} F.
\end{equation}
We calculate the motion of the probability densities by solving this equation using the Verlet algorithm with a time step of $\Delta t = 50$ fs. 
The gradient of the free energy, $\nabla_{\boldsymbol{X}_{i}}F$, can be represented as follows, with the derivation given in Supplemental Material S2:
\begin{eqnarray}
 -\nabla_{\boldsymbol{X}_{i}} F 
 &=&
 -\frac{\partial F}{\partial \boldsymbol{X}_{i}} \\ \nonumber
 &=&
 -\mathbb{E}_{u_k,\boldsymbol{\eta}_k}
 \Big[
 \nabla_{\boldsymbol{r}_{i}}\phi
 \big(
 \bm{1}_{\{u_{1}\leq c_{1}\}},\ldots,
 \bm{1}_{\{u_{N}\leq c_{N}\}},
 \boldsymbol{X}_{1}+\bm{L}_{1}\boldsymbol{\eta}_{1},\ldots,
 \boldsymbol{X}_{N}+\bm{L}_{N}\boldsymbol{\eta}_{N}
 \big)
 \Big].
\end{eqnarray}
Thus, the negative gradient of the free energy, i.e., the mean force, is given by the expectation value of the force derived from the interatomic potential. 
For the mobility, we employed $\kappa=5.56\times10^{-8}$ m$^{2}$/(V$\cdot$ s), which is the mobility of Cu$^{2+}$ in water at 298 K\mbox{\cite{Cumobility,Ion_mobility_equation}}.

To solve the time evolution of occupation probabilities, we calculated the site-resolved chemical potential at atomic site $i$ as follows (Supplemental Material S2):
\begin{eqnarray}
\mu_i
&=&
\frac{\partial F}{\partial c_{i}}\\ \nonumber
&=&\frac{1}{2}k_{B}T  \ln \left(\frac{\Lambda_{i,\rm{Ag}}^{6}}{\Lambda_{i,\rm{Cu}}^{6}} \right)  \\ \nonumber
&+&
\mathbb{E}_{u_{k\neq i}\sim U(0,1),\,\boldsymbol{\eta}_{k}\sim 
N(\boldsymbol{0},\mathbf{I})}
\Big[
\phi \big(
\bm{1}_{\{u_{1}\leq c_{1}\}},\ldots,1,\ldots,
\bm{1}_{\{u_{N}\leq c_{N}\}},
\boldsymbol{X}_{1}+\boldsymbol{L}_{1}\boldsymbol{\eta}_{1},\ldots,
\boldsymbol{X}_{N}+\boldsymbol{L}_{N}\boldsymbol{\eta}_{N}
\big)
\Big] \\ \nonumber
&-&
\mathbb{E}_{u_{k\neq i}\sim U(0,1),\,\boldsymbol{\eta}_{k}\sim 
N(\boldsymbol{0},\mathbf{I})}
\Big[
\phi \big(
\bm{1}_{\{u_{1}\leq c_{1}\}},\ldots,0,\ldots,
\bm{1}_{\{u_{N}\leq c_{N}\}},
\boldsymbol{X}_{1}+\boldsymbol{L}_{1}\boldsymbol{\eta}_{1},\ldots,
\boldsymbol{X}_{N}+\boldsymbol{L}_{N}\boldsymbol{\eta}_{N}
\big)
\Big] \\ \nonumber
&+& k_{B}T \left[\ln c_{i} - \ln(1-c_{i})\right].
\end{eqnarray}
The second and third terms correspond to the many-body potential energies when site $i$ is occupied by Ag and Cu, respectively, while the occupations of the other sites are sampled according to their occupation probabilities using a sample size of 100.
Using the calculated chemical potentials, we updated $c_{i}$ at each site using an implicit method, as described in Supplemental Information S3. 
We set $M_{ij}$ = 0.1 eV$^{-1}$ $\cdot$ s$^{-1}$ because the mobility associated with occupation exchange is unknown, 
and neighboring sites were defined using a cutoff distance of 6 \mbox{\AA}.

\begin{figure}
  \includegraphics{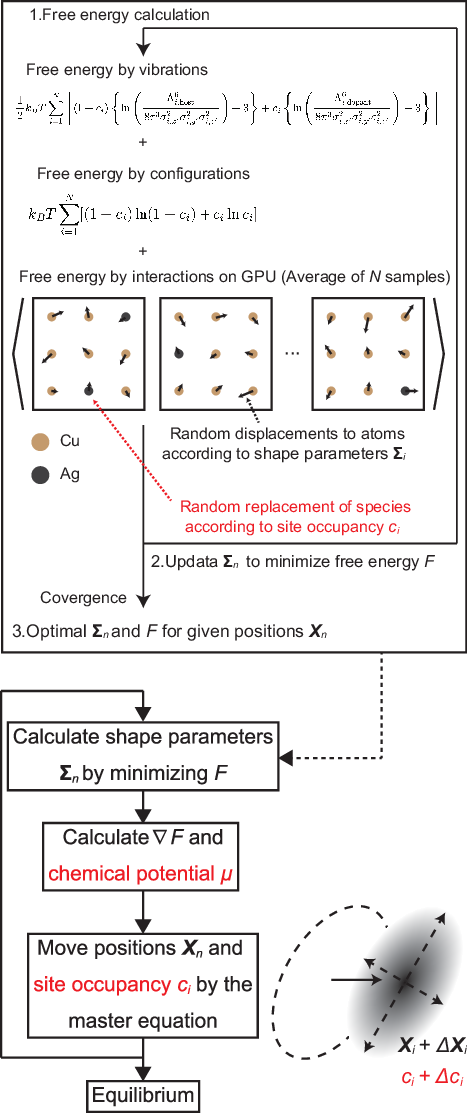}
  \caption{A schematic illustration of the computational workflow for phase-field modeling with dopants, including many-body interactions and fractional site occupancy.}
  \label{fig:temperature}
\end{figure}

\clearpage
\section{Results}
\subsection{Evolution of occupation probability}
Figure 2 shows the evolution of the occupation probability density for $N_{\rm{Ag}}=8$ in the 32-atom system, corresponding to Cu:Ag = 24:8. That is, Ag atoms are initially homogeneously distributed, and each atomic site has an Ag occupation probability of 8/32=0.25. We found that the Ag occupation probability begins to concentrate on specific sublattice positions of the face-centered cubic (fcc) structure, and an ordered-like alloy state is then formed. This behavior is reminiscent of the ordering tendency expected from cluster expansion and observed in Cu–Au systems\mbox{\cite{doi:10.1016/j.actamat.2016.11.048,doi:10.1088/0953-8984/19/8/086201,doi:10.1007/BF02893155}}. Note that, in real bulk systems, Cu and Ag are thermodynamically prone to phase separation\mbox{\cite{PhysRevB.57.6427,doi:10.1007/BF02652162}}; therefore, the ordered-like state obtained here should be interpreted as a finite-size-stabilized occupancy pattern in the small cell. Nevertheless, the result demonstrates that the present modeling framework can explore possible low-energy occupancy patterns.


\clearpage

\begin{figure}
  \includegraphics{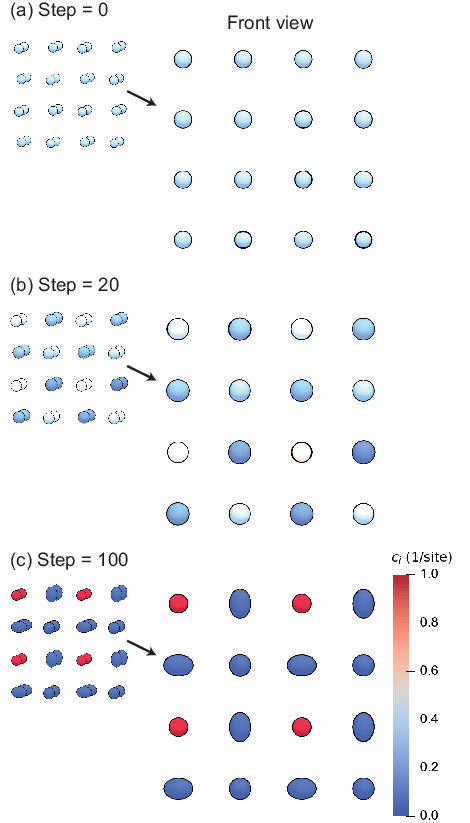}
\caption{Structure evolution of the site occupation probability $c_i$ for $N_{\rm Ag}=8$ in a 32-atom bulk Cu--Ag system (Cu:Ag = 24:8) at (a) 0, (b) 20, and (c) 100 minimization steps. Blue atoms indicate Cu-rich sites, while red atoms indicate sites with high Ag occupation probability.}
  \label{fig:temperature}
\end{figure}

\clearpage
\subsection{Temperature dependence of lattice constants}
To examine the temperature dependence of the lattice constants, we calculate the pressure as follows (the derivation is given in Supplemental Material S2):
\begin{eqnarray}
 P&=&-\Big(\frac{\partial F}{\partial V}\Big)_{T,N}\\ \nonumber
&=&
\mathbb{E}_{u_k\sim U(0,1),\,\bm{\eta}_k\sim N(\boldsymbol{0},\mathbf{I})}
\bigg[
\frac{\partial \phi}{\partial V} \big(
\bm{1}_{\{u_{1}\leq c_{1}\}},\ldots,\bm{1}_{\{u_{N}\leq c_{N}\}},
\boldsymbol{X}_{1}+\boldsymbol{L}_{1}\bm{\eta}_{1},\ldots,
\boldsymbol{X}_{N}+\boldsymbol{L}_{N}\bm{\eta}_{N}
\big)
\bigg].
\end{eqnarray}
Using this pressure, we change the cell volume according to the following equation:
\begin{eqnarray}
\frac{dV}{dt}=M(P-P_{0}),
\end{eqnarray}
where $P_{0}$ is the applied external pressure and $M$ is the volume mobility. In other words, we scale the atomic positions and the volume at each time step as follows:
\begin{eqnarray}
\boldsymbol{X}_{i} &\rightarrow& \mu \boldsymbol{X}_{i}, \nonumber \\
V &\rightarrow& \mu^{3} V,
\end{eqnarray}
where $\mu$ is the scale factor given by
\begin{equation}
\mu=\{1+M' \Delta t (P-P_{0})\}^{\frac{1}{3}},
\end{equation}
and $M' = M/V$ is the mobility per unit volume. In this study, we set $M' = 10.0$~Pa$^{-1}\cdot$s$^{-1}$. Using the above pressure and volume control scheme, we calculated the equilibrium lattice constants by phase-field simulations.

Figure 3 shows the temperature dependence of the lattice constant for different numbers of dopant atoms. 
As the initial condition for the phase-field simulations, Ag atoms are homogeneously distributed in the same manner as in Fig.~2.
For comparison, we calculated the variation of the lattice constant using MD simulations, in which Ag atoms are arranged in the ordered L1$_2$ structure as shown schematically.
We found that the lattice constants calculated by the phase-field simulations are in good agreement with those obtained from the MD simulations.
These results support the validity of the free-energy functional constructed in this study for describing how the lattice constant depends on temperature and dopant concentration, as well as for identifying low-energy configurations.

\clearpage

\begin{figure}
  \includegraphics{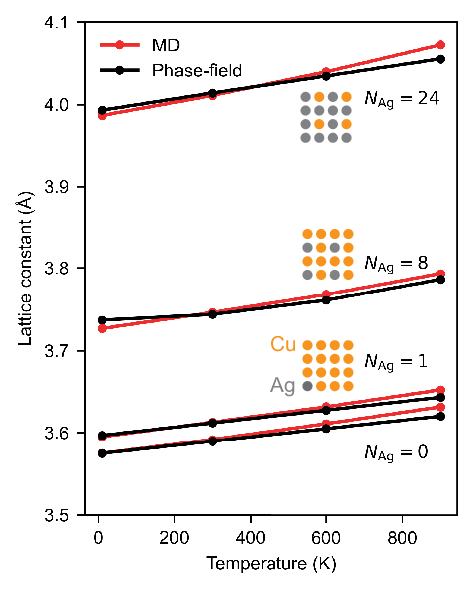}
  \caption{Temperature dependence of the lattice constants of 32-atom Cu–Ag systems obtained from phase-field and MD simulations with different numbers of Ag dopant atoms. The initial Ag distribution is homogeneous in the phase-field simulations, whereas the ordered L1$_2$ structure is used as the initial configuration in the MD simulations, as shown schematically.}
  \label{fig:temperature}
\end{figure}


\clearpage

\subsection{Spatial pattern of dopant segregation at a grain boundary}
Next, we apply the developed methodology to predict the segregation pattern around a grain boundary. Figure 4 shows the time evolution of the Ag occupation probability around the $\Sigma5(310)[001]$ Cu grain boundary doped with $N_{\rm Ag}=4$ in a 76-atom system under the NPT ensemble at $T=1200$ K and $P_{0}=0$ GPa. 
The initial Ag occupation probability is homogeneously distributed over all atomic sites (Fig. 4(a)). As the minimization proceeds, it becomes concentrated at the grain-boundary vertex sites (Figs. 4(b) and 4(c)). 
This tendency is consistent with previous experimental observations and simulations\mbox{\cite{PhysRevLett.110.255502,doi:10.1038/nmat1191}}, 
which have shown that atoms larger than Cu tend to segregate to the vertex regions of grain boundaries because these regions provide larger free volume. 
This demonstrates the ability of our modeling framework to explore low-energy segregation patterns around Cu grain boundaries.

\begin{figure}
  \includegraphics{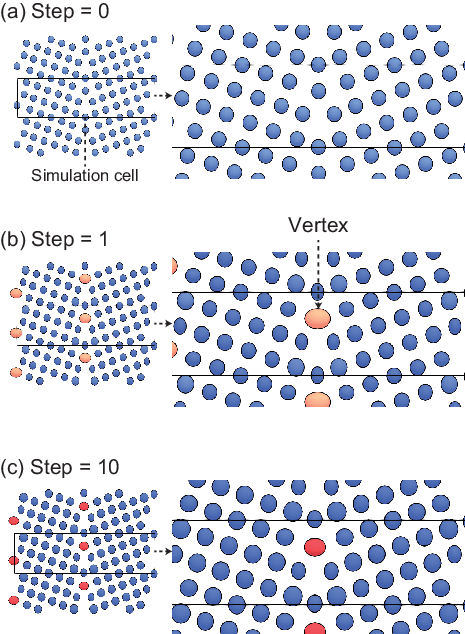}
\caption{Evolution of the Ag occupation probability around a $\Sigma5(310)[001]$ Cu grain boundary doped with $N_{\rm Ag}=4$ in a 76-atom system under the NPT ensemble at $T=1200$ K and $P_{0}=0$ GPa, as calculated using phase-field simulations. Probability distributions are shown at (a) 0, (b) 10, and (c) 100 minimization steps. The color scale is the same as in Fig. 2.}
  \label{fig:temperature}
\end{figure}

\clearpage

We then investigate the variation of the segregation pattern with the number of dopants at room temperature. Figure~5 shows the Ag occupation probability for various numbers of Ag atoms, $N_{\rm Ag}=4$, 8, and 16, under the NPT ensemble at $T=300$ K and $P_{0}=0$ GPa. 
The probability distributions obtained at 1200 K for each number of dopants were used as the initial conditions, assuming that segregation patterns formed at high temperature can be frozen in during cooling. 
We found that Ag atoms preferentially segregate at the vertices of the triangular motif, indicating that these vertices are the primary segregation sites (Fig. 5(a)). 
As the number of dopants increases, additional Ag atoms segregate to sites adjacent to the vertices (Fig. 5(b)). Further excess Ag atoms, which cannot be accommodated at these secondary preferred sites, then segregate at the bottom sites of the triangular motif. 
These results clearly demonstrate that our modeling framework enables the identification of primary, secondary, and tertiary preferred sites for dopant segregation, thereby visualizing atomic-scale pattern formation.

\clearpage


\begin{figure}
  \includegraphics{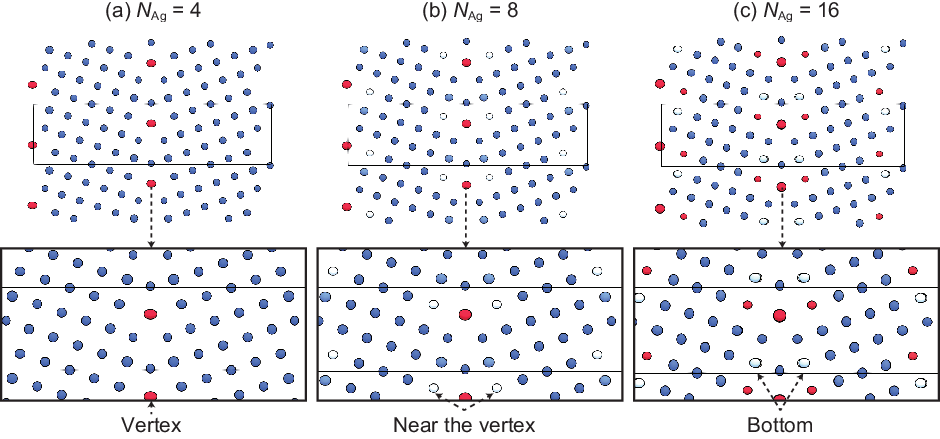}
\caption{Segregation patterns around a $\Sigma5(310)[001]$ Cu grain boundary with various numbers of Ag dopant atoms in a 76-atom system under the NPT ensemble at $T=300$ K and $P_{0}=0$ GPa, as calculated using phase-field simulations. The probability distributions are shown for (a) $N_{\rm Ag}=4$, (b) $N_{\rm Ag}=8$, and (c) $N_{\rm Ag}=16$. The color scale is the same as in Fig. 2.}
  \label{fig:temperature}
\end{figure}


\clearpage

\clearpage

Thus, our modeling successfully explores segregation patterns around grain boundaries driven by chemical potential. We also note several limitations of the present study. (1) We do not consider complexion transitions, which may involve variations in the number of atomic sites. (2) We do not consider correlations between atomic vibrations. This approximation may be reasonable for heavy-atom systems such as metals; however, correlations will become important in systems such as oxides. (3) The calculation of the chemical potential, Eq. (18), requires evaluating the difference in potential energy for each site, which will increase the computational cost for larger systems and multicomponent systems.
 

\clearpage

\clearpage
\section{Conclusion}
In summary, we have developed an atomic-scale phase-field modeling framework that incorporates fractional site occupancy and many-body interactions. Our modeling successfully reproduces the temperature dependence of the lattice constant of Ag-doped bulk Cu. Furthermore, we applied the developed methodology to predict Ag segregation patterns around a Cu grain boundary. We found that Ag atoms first segregate at the vertices of the triangular motif. With increasing dopant concentration, excess Ag atoms occupy sites adjacent to the vertices and subsequently segregate to the bottom sites of the triangular motif. These results demonstrate that our modeling framework extends the phase-field concept to discrete atomic systems and enables the identification of preferential dopant-segregation sites, thereby visualizing atomic-scale pattern formation.


\begin{acknowledgement}
This work was supported by JSPS KAKENHI grant numbers 25K23433.

\end{acknowledgement}

\begin{suppinfo}
The Supporting Information is available free of charge at
\begin{itemize}
  \item PDF: Variational free energy with fractional site occupancy, Derivatives of free energy, and Implicit method for site-occupancy dynamics.
\end{itemize}

\end{suppinfo}\textbf{}

\bibliography{manuscript}

@article{LQchenReview,
author = {Chen, L. Q.},
title = {Phase-Field Models for Microstructure Evolution},
journal = {Annu. Rev. Mater. Res.},
volume = {32},
number = {1},
pages = {113-140},
year = {2002},
doi = {10.1146/annurev.matsci.32.112001.132041},
}

@article{SolidificationReview,
author = {Boettinger, W. J. and Warren, J. A. and Beckermann, C. and Karma, A.},
title = {Phase-Field Simulation of Solidification},
journal = {Annu. Rev. Mater. Res.},
volume = {32},
number = {1},
pages = {163-194},
year = {2002}
}

@Article{Ambati2015,
author={Ambati, Marreddy
and Gerasimov, Tymofiy
and De Lorenzis, Laura},
title={A Review on Phase-Field Models of Brittle Fracture and a New Fast Hybrid Formulation},
journal={Comput. Mech.},
year={2015},
month={Feb},
day={01},
volume={55},
number={2},
pages={383-405},
}

@article{doi:10.1098/rsta.2015.0166,
author = {Beyerlein, I. J.  and Hunter, A. },
title = {Understanding dislocation mechanics at the mesoscale using phase field dislocation dynamics},
journal = {Phil. Trans. R. Soc. A},
volume = {374},
number = {2066},
pages = {20150166},
year = {2016},
doi = {10.1098/rsta.2015.0166},
URL = {https://royalsocietypublishing.org/doi/abs/10.1098/rsta.2015.0166},
eprint = {https://royalsocietypublishing.org/doi/pdf/10.1098/rsta.2015.0166}
}

@article{doi:10.1080/14786444908561372,
author = {A. F. Devonshire },
title = {{XCVI.} {Theory} of Barium Titanate},
journal = {Phil. Mag.},
volume = {40},
number = {309},
pages = {1040-1063},
year  = {1949},
publisher = {Taylor & Francis},
doi = {10.1080/14786444908561372},
}

@article{doi:10.1080/14786445108561354,
author = {A.F. Devonshire},
title = {{CIX.} {Theory} of barium titanate-{Part II} },
journal = {Phil. Mag.},
volume = {42},
number = {333},
pages = {1065--1079},
year = {1951},
publisher = {Taylor \& Francis},
doi = {10.1080/14786445108561354},

}

@article{LQchen_review_thin_film,
author = {Chen, L. Q.},
title = {Phase-Field Method of Phase Transitions/Domain Structures in Ferroelectric Thin Films: A Review},
journal = {J. Am. Ceram. Soc.},
volume = {91},
number = {6},
pages = {1835-1844},
doi = {https://doi.org/10.1111/j.1551-2916.2008.02413.x},
url = {https://ceramics.onlinelibrary.wiley.com/doi/abs/10.1111/j.1551-2916.2008.02413.x},
eprint = {https://ceramics.onlinelibrary.wiley.com/doi/pdf/10.1111/j.1551-2916.2008.02413.x},
year = {2008}
}

@Article{doi:10.1038/nature16463,
author={Yadav, A. K.
and Nelson, C. T.
and Hsu, S. L.
and Hong, Z.
and Clarkson, J. D.
and Schlep{\"u}tz, C. M.
and Damodaran, A. R.
and Shafer, P.
and Arenholz, E.
and Dedon, L. R.
and Chen, D.
and Vishwanath, A.
and Minor, A. M.
and Chen, L. Q.
and Scott, J. F.
and Martin, L. W.
and Ramesh, R.},
title={Observation of Polar Vortices in Oxide Superlattices},
journal={Nature},
year={2016},
month={Feb},
day={01},
volume={530},
number={7589},
pages={198-201},
issn={1476-4687},
doi={10.1038/nature16463},
url={https://doi.org/10.1038/nature16463}
}

@Article{doi:10.1038/s41586-019-1092-8,
author={Das, S.
and Tang, Y. L.
and Hong, Z.
and Gon{\c{c}}alves, M. A. P.
and McCarter, M. R.
and Klewe, C.
and Nguyen, K. X.
and G{\'o}mez-Ortiz, F.
and Shafer, P.
and Arenholz, E.
and Stoica, V. A.
and Hsu, S.-L.
and Wang, B.
and Ophus, C.
and Liu, J. F.
and Nelson, C. T.
and Saremi, S.
and Prasad, B.
and Mei, A. B.
and Schlom, D. G.
and {\'I}{\~{n}}iguez, J.
and Garc{\'i}a-Fern{\'a}ndez, P.
and Muller, D. A.
and Chen, L. Q.
and Junquera, J.
and Martin, L. W.
and Ramesh, R.},
title={Observation of room-temperature polar skyrmions},
journal={Nature},
year={2019},
month={Apr},
day={01},
volume={568},
number={7752},
pages={368-372},
issn={1476-4687},
doi={10.1038/s41586-019-1092-8},
url={https://doi.org/10.1038/s41586-019-1092-8}
}

@Article{doi:10.1038/s41563-020-0694-8,
author={Wang, Y. J.
and Feng, Y. P.
and Zhu, Y. L.
and Tang, Y. L.
and Yang, L. X.
and Zou, M. J.
and Geng, W. R.
and Han, M. J.
and Guo, X. W.
and Wu, B.
and Ma, X. L.},
title={Polar meron lattice in strained oxide ferroelectrics},
journal={Nat. Mater.},
year={2020},
month={Aug},
day={01},
volume={19},
number={8},
pages={881-886},
issn={1476-4660},
doi={10.1038/s41563-020-0694-8},
url={https://doi.org/10.1038/s41563-020-0694-8}
}

@Article{doi:10.1038/s41467-023-36950-x,
author={Shao, Yu-Tsun
and Das, Sujit
and Hong, Zijian
and Xu, Ruijuan
and Chandrika, Swathi
and G{\'o}mez-Ortiz, Fernando
and Garc{\'i}a-Fern{\'a}ndez, Pablo
and Chen, Long-Qing
and Hwang, Harold Y.
and Junquera, Javier
and Martin, Lane W.
and Ramesh, Ramamoorthy
and Muller, David A.},
title={Emergent chirality in a polar meron to skyrmion phase transition},
journal={Nat. Commun.},
year={2023},
month={Mar},
day={13},
volume={14},
number={1},
pages={1355},
issn={2041-1723},
doi={10.1038/s41467-023-36950-x},
url={https://doi.org/10.1038/s41467-023-36950-x}
}

@article{doi:10.1126/science.1259869,
author = {Y. L. Tang  and Y. L. Zhu  and X. L. Ma  and A. Y. Borisevich  and A. N. Morozovska  and E. A. Eliseev  and W. Y. Wang  and Y. J. Wang  and Y. B. Xu  and Z. D. Zhang  and S. J. Pennycook },
title = {Observation of a Periodic Array of Flux-Closure Quadrants in Strained Ferroelectric {PbTiO}$_3$ films},
journal = {Science},
volume = {348},
number = {6234},
pages = {547-551},
year = {2015},
doi = {10.1126/science.1259869},
}

@article{CHOUDHURY20055313,
title = {Phase-field simulation of polarization switching and domain evolution in ferroelectric polycrystals},
journal = {Acta Mater.},
volume = {53},
number = {20},
pages = {5313-5321},
year = {2005},
issn = {1359-6454},
doi = {https://doi.org/10.1016/j.actamat.2005.07.040},
url = {https://www.sciencedirect.com/science/article/pii/S1359645405004672},
author = {S. Choudhury and Y.L. Li and C.E. Krill and L.-Q. Chen},
}

@Article{Gao2013,
author={Gao, Peng
and Britson, Jason
and Jokisaari, Jacob R.
and Nelson, Christopher T.
and Baek, Seung-Hyub
and Wang, Yiran
and Eom, Chang-Beom
and Chen, Long-Qing
and Pan, Xiaoqing},
title={Atomic-scale mechanisms of ferroelastic domain-wall-mediated ferroelectric switching},
journal={Nat. Commun.},
year={2013},
month={Nov},
day={21},
volume={4},
number={1},
pages={2791},
issn={2041-1723},
doi={10.1038/ncomms3791},
url={https://doi.org/10.1038/ncomms3791}
}

@Article{Park2018,
author={Park, Sung Min
and Wang, Bo
and Das, Saikat
and Chae, Seung Chul
and Chung, Jin-Seok
and Yoon, Jong-Gul
and Chen, Long-Qing
and Yang, Sang Mo
and Noh, Tae Won},
title={Selective control of multiple ferroelectric switching pathways using a trailing flexoelectric field},
journal={Nat. Nanotechnol.},
year={2018},
month={May},
day={01},
volume={13},
number={5},
pages={366-370},
issn={1748-3395},
doi={10.1038/s41565-018-0083-5},
url={https://doi.org/10.1038/s41565-018-0083-5}
}

@article{WANG2021117383,
title = {The rectilinear motion of the individual asymmetrical skyrmion driven by temperature gradients},
journal = {Acta Mater.},
volume = {221},
pages = {117383},
year = {2021},
issn = {1359-6454},
doi = {https://doi.org/10.1016/j.actamat.2021.117383},
url = {https://www.sciencedirect.com/science/article/pii/S135964542100762X},
author = {Yu Wang and Takahiro Shimada and Jie Wang and Takayuki Kitamura and Hiroyuki Hirakata},
}

@article{NatureYang2025,
author={Yang, Bingbing
and Liu, Yiqian
and Jiang, Ru-Jian
and Lan, Shun
and Liu, Su-Zhen
and Zhou, Zhifang
and Dou, Lvye
and Zhang, Min
and Huang, Houbing
and Chen, Long-Qing
and Zhu, Yin-Lian
and Zhang, Shujun
and Ma, Xiu-Liang
and Nan, Ce-Wen
and Lin, Yuan-Hua},
title={Enhanced energy storage in antiferroelectrics via antipolar frustration},
journal={Nature},
year={2025},
month={Jan},
day={01},
volume={637},
number={8048},
pages={1104-1110},
issn={1476-4687},
doi={10.1038/s41586-024-08505-7},
url={https://doi.org/10.1038/s41586-024-08505-7}
}

@article{
doi:10.1126/science.abb3209,
author = {Huajun Liu  and Haijun Wu  and Khuong Phuong Ong  and Tiannan Yang  and Ping Yang  and Pranab Kumar Das  and Xiao Chi  and Yang Zhang  and Caozheng Diao  and Wai Kong Alaric Wong  and Eh Piew Chew  and Yi Fan Chen  and Chee Kiang Ivan Tan  and Andrivo Rusydi  and Mark B. H. Breese  and David J. Singh  and Long-Qing Chen  and Stephen J. Pennycook  and Kui Yao },
title = {Giant piezoelectricity in oxide thin films with nanopillar structure},
journal = {Science},
volume = {369},
number = {6501},
pages = {292-297},
year = {2020},
doi = {10.1126/science.abb3209},
URL = {https://www.science.org/doi/abs/10.1126/science.abb3209},
eprint = {https://www.science.org/doi/pdf/10.1126/science.abb3209},
}

@article{doi:10.1126/science.adq4147,
author = {Vivek Devulapalli  and Enze Chen  and Tobias Brink  and Timofey Frolov  and Christian H. Liebscher },
title = {Topological grain boundary segregation transitions},
journal = {Science},
volume = {386},
number = {6720},
pages = {420-424},
year = {2024},
doi = {10.1126/science.adq4147},
URL = {https://www.science.org/doi/abs/10.1126/science.adq4147},
eprint = {https://www.science.org/doi/pdf/10.1126/science.adq4147},
}

@Article{doi:10.1038/s41586-020-2082-6,
author={Meiners, Thorsten
and Frolov, Timofey
and Rudd, Robert E.
and Dehm, Gerhard
and Liebscher, Christian H.},
title={Observations of grain-boundary phase transformations in an elemental metal},
journal={Nature},
year={2020},
month={Mar},
day={01},
volume={579},
number={7799},
pages={375-378},
issn={1476-4687},
doi={10.1038/s41586-020-2082-6},
url={https://doi.org/10.1038/s41586-020-2082-6}
}

@article{PhysRevLett.120.267601,
  title = {Atomic-scale measurement of flexoelectric polarization at ${\mathrm{SrTiO}}_{3}$ dislocations},
  author = {Gao, Peng and Yang, Shuzhen and Ishikawa, Ryo and Li, Ning and Feng, Bin and Kumamoto, Akihito and Shibata, Naoya and Yu, Pu and Ikuhara, Yuichi},
  journal = {Phys. Rev. Lett.},
  volume = {120},
  issue = {26},
  pages = {267601},
  numpages = {6},
  year = {2018},
  month = {Jun},
  publisher = {American Physical Society},
  doi = {10.1103/PhysRevLett.120.267601},
  url = {https://link.aps.org/doi/10.1103/PhysRevLett.120.267601}
}

@article{doi:10.1126/science.adh7670,
author = {Sebastian Calderon  and John Hayden  and Steven M. Baksa  and William Tzou  and Susan Trolier-McKinstry  and Ismaila Dabo  and Jon-Paul Maria  and Elizabeth C. Dickey },
title = {Atomic-scale polarization switching in wurtzite ferroelectrics},
journal = {Science},
volume = {380},
number = {6649},
pages = {1034-1038},
year = {2023},
doi = {10.1126/science.adh7670},
URL = {https://www.science.org/doi/abs/10.1126/science.adh7670},
eprint = {https://www.science.org/doi/pdf/10.1126/science.adh7670},
}

@Article{doi:0.1038/s41467-021-21019-4,
author={Furukawa, Shunsuke
and Wu, Jianyun
and Koyama, Masaya
and Hayashi, Keisuke
and Hoshino, Norihisa
and Takeda, Takashi
and Suzuki, Yasutaka
and Kawamata, Jun
and Saito, Masaichi
and Akutagawa, Tomoyuki},
title={Ferroelectric columnar assemblies from the bowl-to-bowl inversion of aromatic cores},
journal={Nat. Commun.},
year={2021},
month={Feb},
day={03},
volume={12},
number={1},
pages={768},
issn={2041-1723},
doi={10.1038/s41467-021-21019-4},
url={https://doi.org/10.1038/s41467-021-21019-4}
}

@Article{doi:10.1038/36069,
author={Bune, A. V.
and Fridkin, V. M.
and Ducharme, Stephen
and Blinov, L. M.
and Palto, S. P.
and Sorokin, A. V.
and Yudin, S. G.
and Zlatkin, A.},
title={Two-dimensional ferroelectric films},
journal={Nature},
year={1998},
month={Feb},
day={01},
volume={391},
number={6670},
pages={874-877},
issn={1476-4687},
doi={10.1038/36069},
url={https://doi.org/10.1038/36069}
}

@article{doi:10.1103/PhysRevE.70.051605,
  title = {Modeling elastic and plastic deformations in nonequilibrium processing using phase field crystals},
  author = {Elder, K. R. and Grant, Martin},
  journal = {Phys. Rev. E},
  volume = {70},
  issue = {5},
  pages = {051605},
  numpages = {18},
  year = {2004},
  month = {Nov},
  publisher = {American Physical Society},
  doi = {10.1103/PhysRevE.70.051605},
  url = {https://link.aps.org/doi/10.1103/PhysRevE.70.051605}
}

@article{doi:10.1103/PhysRevE.79.035701,
  title = {Molecular dynamics on diffusive time scales from the phase-field-crystal equation},
  author = {Chan, Pak Yuen and Goldenfeld, Nigel and Dantzig, Jon},
  journal = {Phys. Rev. E},
  volume = {79},
  issue = {3},
  pages = {035701},
  numpages = {4},
  year = {2009},
  month = {Mar},
  publisher = {American Physical Society},
  doi = {10.1103/PhysRevE.79.035701},
  url = {https://link.aps.org/doi/10.1103/PhysRevE.79.035701}
}

@article{doi:10.1103/PhysRevE.73.031609,
  title = {Diffusive atomistic dynamics of edge dislocations in two dimensions},
  author = {Berry, J. and Grant, M. and Elder, K. R.},
  journal = {Phys. Rev. E},
  volume = {73},
  issue = {3},
  pages = {031609},
  numpages = {12},
  year = {2006},
  month = {Mar},
  publisher = {American Physical Society},
  doi = {10.1103/PhysRevE.73.031609},
  url = {https://link.aps.org/doi/10.1103/PhysRevE.73.031609}
}

@Article{doi:10.1007/s11837-007-0095-3,
author={Provatas, N.
and Dantzig, J. A.
and Athreya, B.
and Chan, P.
and Stefanovic, P.
and Goldenfeld, N.
and Elder, K. R.},
title={Using the phase-field crystal method in the multi-scale modeling of microstructure evolution},
journal={JOM},
year={2007},
month={Jul},
day={01},
volume={59},
number={7},
pages={83-90}
}

@article{doi:10.1103/PhysRevMaterials.4.013802,
  title = {Displacive phase-field crystal model},
  author = {Alster, Eli and Elder, K. R. and Voorhees, Peter W.},
  journal = {Phys. Rev. Mater.},
  volume = {4},
  issue = {1},
  pages = {013802},
  numpages = {13},
  year = {2020},
  month = {Jan},
  publisher = {American Physical Society},
  doi = {10.1103/PhysRevMaterials.4.013802},
}

@article{PhysRevLett.63.624,
  title = {Finite-Temperature Defect Properties from Free-Energy Minimization},
  author = {LeSar, R. and Najafabadi, R. and Srolovitz, D. J.},
  journal = {Phys. Rev. Lett.},
  volume = {63},
  issue = {6},
  pages = {624--627},
  numpages = {0},
  year = {1989},
  month = {Aug},
}

@article{doi:10.1063/1.460547,
author = {LeSar,R.  and Najafabadi,R.  and Srolovitz,D. J. },
title = {Thermodynamics of Solid and Liquid Embedded‐Atom‐Method Metals: A Variational Study},
journal = {J. Chem. Phys.},
volume = {94},
number = {7},
pages = {5090-5097},
year = {1991},
doi = {10.1063/1.460547}
}

@article{PhysRevB.84.054103,
  title = {Diffusive Molecular Dynamics and Its Application to Nanoindentation and Sintering},
  author = {Li, Ju and Sarkar, Sanket and Cox, William T. and Lenosky, Thomas J. and Bitzek, Erik and Wang, Yunzhi},
  journal = {Phys. Rev. B},
  volume = {84},
  issue = {5},
  pages = {054103},
  numpages = {8},
  year = {2011},
  month = {Aug},
  publisher = {American Physical Society},
  doi = {10.1103/PhysRevB.84.054103},
  url = {https://link.aps.org/doi/10.1103/PhysRevB.84.054103}
}

@article{PhysRevB.110.104107,
  title = {Phase field modeling for atoms},
  author = {Masuda, Kairi and Rappe, Andrew M.},
  journal = {Phys. Rev. B},
  volume = {110},
  issue = {10},
  pages = {104107},
  numpages = {9},
  year = {2024},
  month = {Sep},
  publisher = {American Physical Society},
  doi = {10.1103/PhysRevB.110.104107},
  url = {https://link.aps.org/doi/10.1103/PhysRevB.110.104107}
}

@article{doi:10.1103/ngky-crt4,
  title = {Atomic-scale phase-field modeling for ferroelectrics},
  author = {Masuda, Kairi and Rappe, Andrew M.},
  journal = {Phys. Rev. B},
  volume = {112},
  issue = {5},
  pages = {054107},
  numpages = {10},
  year = {2025},
  month = {Aug},
  publisher = {American Physical Society},
  doi = {10.1103/ngky-crt4},
  url = {https://link.aps.org/doi/10.1103/ngky-crt4}
}

@Article{doi:10.1038/s41467-024-49437-0,
author={Hunnestad, Kasper A.
and Das, Hena
and Hatzoglou, Constantinos
and Holtz, Megan
and Brooks, Charles M.
and van Helvoort, Antonius T. J.
and Muller, David A.
and Schlom, Darrell G.
and Mundy, Julia A.
and Meier, Dennis},
title={{3D} oxygen vacancy distribution and defect-property relations in an oxide heterostructure},
journal={Nat. Commun.},
year={2024},
month={Jun},
day={26},
volume={15},
number={1},
pages={5400},
issn={2041-1723},
doi={10.1038/s41467-024-49437-0},
url={https://doi.org/10.1038/s41467-024-49437-0}
}

@Article{doi:10.1038/s41467-022-30018-y,
author={Su, Lei
and Huyan, Huaixun
and Sarkar, Abhishek
and Gao, Wenpei
and Yan, Xingxu
and Addiego, Christopher
and Kruk, Robert
and Hahn, Horst
and Pan, Xiaoqing},
title={Direct observation of elemental fluctuation and oxygen octahedral distortion-dependent charge distribution in high entropy oxides},
journal={Nat. Commun.},
year={2022},
month={Apr},
day={29},
volume={13},
number={1},
pages={2358},
issn={2041-1723},
doi={10.1038/s41467-022-30018-y},
url={https://doi.org/10.1038/s41467-022-30018-y}
}

@Article{doi:10.1038/ncomms11079,
author={Feng, Bin
and Yokoi, Tatsuya
and Kumamoto, Akihito
and Yoshiya, Masato
and Ikuhara, Yuichi
and Shibata, Naoya},
title={Atomically ordered solute segregation behaviour in an oxide grain boundary},
journal={Nat. Commun.},
year={2016},
month={Mar},
day={23},
volume={7},
number={1},
pages={11079},
issn={2041-1723},
doi={10.1038/ncomms11079},
url={https://doi.org/10.1038/ncomms11079}
}

@Article{Batatia2025,
author={Batatia, Ilyes
and Batzner, Simon
and Kov{\'a}cs, D{\'a}vid P{\'e}ter
and Musaelian, Albert
and Simm, Gregor N. C.
and Drautz, Ralf
and Ortner, Christoph
and Kozinsky, Boris
and Cs{\'a}nyi, G{\'a}bor},
title={The design space of E(3)-equivariant atom-centred interatomic potentials},
journal={Nat. Mach. Intell.},
year={2025},
month={Jan},
day={01},
volume={7},
number={1},
pages={56-67},
issn={2522-5839},
doi={10.1038/s42256-024-00956-x},
url={https://doi.org/10.1038/s42256-024-00956-x}
}

@article{PhysRevB.89.064302,
  title = {Anharmonic free energies and phonon dispersions from the stochastic self-consistent harmonic approximation: Application to platinum and palladium hydrides},
  author = {Errea, Ion and Calandra, Matteo and Mauri, Francesco},
  journal = {Phys. Rev. B},
  volume = {89},
  issue = {6},
  pages = {064302},
  numpages = {16},
  year = {2014},
  month = {Feb},
  publisher = {American Physical Society},
  doi = {10.1103/PhysRevB.89.064302},
  url = {https://link.aps.org/doi/10.1103/PhysRevB.89.064302}
}

@article{PhysRevB.98.024106,
  title = {Pressure and stress tensor of complex anharmonic crystals within the stochastic self-consistent harmonic approximation},
  author = {Monacelli, Lorenzo and Errea, Ion and Calandra, Matteo and Mauri, Francesco},
  journal = {Phys. Rev. B},
  volume = {98},
  issue = {2},
  pages = {024106},
  numpages = {17},
  year = {2018},
  month = {Jul},
  publisher = {American Physical Society},
  doi = {10.1103/PhysRevB.98.024106},
  url = {https://link.aps.org/doi/10.1103/PhysRevB.98.024106}
}

@Article{doi:10.1038/s41586-020-1955-z,
author={Errea, Ion
and Belli, Francesco
and Monacelli, Lorenzo
and Sanna, Antonio
and Koretsune, Takashi
and Tadano, Terumasa
and Bianco, Raffaello
and Calandra, Matteo
and Arita, Ryotaro
and Mauri, Francesco
and Flores-Livas, Jos{\'e} A.},
title={Quantum crystal structure in the 250-kelvin superconducting lanthanum hydride},
journal={Nature},
year={2020},
month={Feb},
day={01},
volume={578},
number={7793},
pages={66-69},
issn={1476-4687},
doi={10.1038/s41586-020-1955-z},
url={https://doi.org/10.1038/s41586-020-1955-z}
}

@article{doi:10.1088/1361-648X/ac066b,
doi = {10.1088/1361-648X/ac066b},
url = {https://doi.org/10.1088/1361-648X/ac066b},
year = {2021},
month = {jul},
publisher = {IOP Publishing},
volume = {33},
number = {36},
pages = {363001},
author = {Monacelli, Lorenzo and Bianco, Raffaello and Cherubini, Marco and Calandra, Matteo and Errea, Ion and Mauri, Francesco},
title = {The stochastic self-consistent harmonic approximation: calculating vibrational properties of materials with full quantum and anharmonic effects},
journal = {J. Phys. Condens. Matter.},
}

@article{PhysRevB.110.144101,
  title = {Beyond {Gaussian} fluctuations of quantum anharmonic nuclei: {The} case of rotational degrees of freedom},
  author = {Siciliano, Antonio and Monacelli, Lorenzo and Mauri, Francesco},
  journal = {Phys. Rev. B},
  volume = {110},
  issue = {14},
  pages = {144101},
  numpages = {18},
  year = {2024},
  month = {Oct},
  publisher = {American Physical Society},
  doi = {10.1103/PhysRevB.110.144101},
  url = {https://link.aps.org/doi/10.1103/PhysRevB.110.144101}
}

@article{SOBOL196786,
title = {On the distribution of points in a cube and the approximate evaluation of integrals},
journal = {USSR Comput. Math. Math. Phys.},
volume = {7},
number = {4},
pages = {86-112},
year = {1967},
issn = {0041-5553},
doi = {https://doi.org/10.1016/0041-5553(67)90144-9},
url = {https://www.sciencedirect.com/science/article/pii/0041555367901449},
author = {I.M Sobol'}
}

@inproceedings{paszke2017automatic,
  title={Automatic differentiation in PyTorch},
  author={Paszke, Adam and Gross, Sam and Chintala, Soumith and Chanan, Gregory and Yang, Edward and DeVito, Zachary and Lin, Zeming and Desmaison, Alban and Antiga, Luca and Lerer, Adam},
  booktitle={NIPS-W},
  year={2017}
}

@article{LBFGS,
 ISSN = {00255718, 10886842},
 URL = {http://www.jstor.org/stable/2006193},
 author = {Jorge Nocedal},
 journal = {Math. Comput.},
 number = {151},
 pages = {773--782},
 publisher = {American Mathematical Society},
 title = {Updating Quasi-Newton Matrices with Limited Storage},
 urldate = {2022-12-18},
 volume = {35},
 year = {1980}
}

@ARTICLE{2020SciPy-NMeth,
  author  = {Virtanen, Pauli and Gommers, Ralf and Oliphant, Travis E. and
            Haberland, Matt and Reddy, Tyler and Cournapeau, David and
            Burovski, Evgeni and Peterson, Pearu and Weckesser, Warren and
            Bright, Jonathan and {van der Walt}, St{\'e}fan J. and
            Brett, Matthew and Wilson, Joshua and Millman, K. Jarrod and
            Mayorov, Nikolay and Nelson, Andrew R. J. and Jones, Eric and
            Kern, Robert and Larson, Eric and Carey, C J and
            Polat, {\.I}lhan and Feng, Yu and Moore, Eric W. and
            {VanderPlas}, Jake and Laxalde, Denis and Perktold, Josef and
            Cimrman, Robert and Henriksen, Ian and Quintero, E. A. and
            Harris, Charles R. and Archibald, Anne M. and
            Ribeiro, Ant{\^o}nio H. and Pedregosa, Fabian and
            {van Mulbregt}, Paul and {SciPy 1.0 Contributors}},
  title   = {{{SciPy} 1.0: Fundamental Algorithms for Scientific
            Computing in Python}},
  journal = {Nat. Methods},
  year    = {2020},
  volume  = {17},
  pages   = {261--272},
  adsurl  = {https://rdcu.be/b08Wh},
  doi     = {10.1038/s41592-019-0686-2},
}

@book{Cumobility,
edition = {8th ed.},
language = {eng},
address = {New York, United States},
booktitle = {Quantitative Chemical Analysis},
author = {Daniel C. Harris},
publisher = {W. H. Freeman and Company},
title = {Quantitative Chemical Analysis},
year = {2010},
pages ={314}
}

@book{Ion_mobility_equation,
edition = {3rd ed.},
language = {eng},
address = {New York, United States},
booktitle = {Quantitative Chemical Analysis},
author = {Francis F. Chen},
publisher = {Springer},
title = {Introduction to Plasma Physics and Controlled Fusion},
year = {2016},
pages ={148}
}

@article{doi:10.1016/j.actamat.2016.11.048,
title = {Analysis of short-range order in {Cu}$_3${Au} using {X}-ray pair distribution functions},
journal = {Acta Mater.},
volume = {125},
pages = {15-26},
year = {2017},
issn = {1359-6454},
doi = {https://doi.org/10.1016/j.actamat.2016.11.048},
url = {https://www.sciencedirect.com/science/article/pii/S1359645416309156},
author = {L.R. Owen and H.Y. Playford and H.J. Stone and M.G. Tucker},
}

@article{doi:10.1088/0953-8984/19/8/086201,
doi = {10.1088/0953-8984/19/8/086201},
url = {https://doi.org/10.1088/0953-8984/19/8/086201},
year = {2007},
month = {feb},
publisher = {},
volume = {19},
number = {8},
pages = {086201},
author = {Sanati, Mahdi and Zunger, Alex},
title = {Evolution of $L$ 1$_2$ ordered domains in fcc {Cu}$_3${Au} alloy},
journal = {J. Phys. Condens. Matter.},
}

@Article{doi:10.1007/BF02893155,
author={Okamoto, H.
and Chakrabarti, D. J.
and Laughlin, D. E.
and Massalski, T. B.},
title={The {Au}-{Cu} ({Gold-Copper}) system},
journal={J. Phase Equilib.},
year={1987},
month={Oct},
day={01},
volume={8},
number={5},
pages={454-474},
issn={1054-9714},
doi={10.1007/BF02893155},
url={https://doi.org/10.1007/BF02893155}
}

@article{PhysRevB.57.6427,
  title = {{Cu-Au}, {Ag-Au}, {Cu-Ag}, and {Ni-Au} intermetallics: First-principles study of temperature-composition phase diagrams and structures},
  author = {Ozoli\ifmmode \mbox{\c{n}}\else \c{n}\fi{}\ifmmode \check{s}\else \v{s}\fi{}, V. and Wolverton, C. and Zunger, Alex},
  journal = {Phys. Rev. B},
  volume = {57},
  issue = {11},
  pages = {6427--6443},
  numpages = {0},
  year = {1998},
  month = {Mar},
  publisher = {American Physical Society},
  doi = {10.1103/PhysRevB.57.6427},
  url = {https://link.aps.org/doi/10.1103/PhysRevB.57.6427}
}

@Article{doi:10.1007/BF02652162,
author={Subramanian, P. R.
and Perepezko, J. H.},
title={The {Ag}-{Cu} ({Silver-Copper}) system},
journal={J. Phase Equilib.},
year={1993},
month={Feb},
day={01},
volume={14},
number={1},
pages={62-75},
issn={1054-9714},
doi={10.1007/BF02652162},
url={https://doi.org/10.1007/BF02652162}
}

@article{PhysRevLett.110.255502,
  title = {Effect of interface phase transformations on diffusion and segregation in high-angle grain boundaries},
  author = {Frolov, T. and Divinski, S. V. and Asta, M. and Mishin, Y.},
  journal = {Phys. Rev. Lett.},
  volume = {110},
  issue = {25},
  pages = {255502},
  numpages = {5},
  year = {2013},
  month = {Jun},
  publisher = {American Physical Society},
  doi = {10.1103/PhysRevLett.110.255502},
  url = {https://link.aps.org/doi/10.1103/PhysRevLett.110.255502}
}

@Article{doi:10.1038/nmat1191,
author={Duscher, Gerd
and Chisholm, Matthew F.
and Alber, Uwe
and R{\"u}hle, Manfred},
title={Bismuth-induced embrittlement of copper grain boundaries},
journal={Nat. Mater.},
year={2004},
month={Sep},
day={01},
volume={3},
number={9},
pages={621-626},
issn={1476-4660},
doi={10.1038/nmat1191},
url={https://doi.org/10.1038/nmat1191}
}

@Article{Horiuchi2008,
author={Horiuchi, Sachio
and Tokura, Yoshinori},
title={Organic Ferroelectrics},
journal={Nat. Mater.},
year={2008},
month={May},
day={01},
volume={7},
number={5},
pages={357-366},
}

@article{elastin,
author = {Yuanming Liu  and Hong Ling Cai  and Matthew Zelisko  and Yunjie Wang  and Jinglan Sun  and Fei Yan  and Feiyue Ma  and Peiqi Wang  and Qian Nataly Chen  and Hairong Zheng  and Xiangjian Meng  and Pradeep Sharma  and Yanhang Zhang  and Jiangyu Li },
title = {Ferroelectric Switching of Elastin},
journal = {Proc. Natl. Acad. Sci. U.S.A.},
volume = {111},
number = {27},
pages = {E2780-E2786},
year = {2014},
}

@article{PhysRev.36.823,
  title = {On the Theory of the Brownian Motion},
  author = {Uhlenbeck, G. E. and Ornstein, L. S.},
  journal = {Phys. Rev.},
  volume = {36},
  issue = {5},
  pages = {823--841},
  numpages = {0},
  year = {1930},
  month = {Sep},
  publisher = {American Physical Society},
  doi = {10.1103/PhysRev.36.823},
  url = {https://link.aps.org/doi/10.1103/PhysRev.36.823}
}

\end{document}